\documentstyle[psfig,conf_iap]{article}
\begin{document}
\newcommand{\apjs}{ApJS\,\,}
\heading{Redshift Evolution of the Low-Column Density Ly$\alpha$ Forest} 
\par\medskip\noindent
\author{Esther M. Hu$^1$}
\address{Institute for Astronomy, 2680 Woodlawn Drive, Honolulu, HI 96822, USA}

\begin{abstract}
Knowledge of the evolution of the low column density end of the
Ly$\alpha$ forest ($12.8 < {\rm log}\ N_{\rm HI} < 16.0\ {\rm
cm}^{-2}$) can provide substantial insight for models of structure
formation and studies of the evolution of the ionizing background.
Here we describe the evolution of the $b$ values and column densities
of forest clouds over the redshift range $2<z<4$.  While the
distribution function for clouds with column densities $N_{\rm HI} <
3 \times 10^{14}\ {\rm cm}^{-2}$ can be well described by a single
invariant power law fit with slope $\sim -1.5$ over this range, at
higher column densities a break in this distribution function is
seen, with a deficiency of high column density clouds relative to the
single power-law fit, and with the break point moving to lower column
densities for clouds at lower redshifts.  The median and cut-off
(minimum) $b$ values are seen to decrease at higher redshifts.
\end{abstract}
\section{Introduction}
Recent CDM-based numerical simulations \cite{Zha95},\cite{Dav97},\cite{Zha97}
have shown that the Ly$\alpha$ forest clouds are a natural consequence 
of structure formation in the intergalactic gas, and a general picture 
\cite{Bi97},\cite{Hui97} has emerged within which their properties may be
understood.  In particular, the low column density ($12.8 <\ {\rm
log}\ N_{\rm HI} < 16.0\ {\rm cm}^{-2}$) forest corresponds to
underdense or slightly overdense regions in the IGM, which are easily
modeled or analytically simulated, and which contain a substantial
fraction of the baryonic material in the intergalactic gas.  Studies
of this component over as wide a redshift range as possible can
provide information which: (1) constrains models for structure
formation, (2) provides normalization for studies of
He\thinspace{II}, and (3) addresses the evolution of the shape and
normalization of the ionizing background radiation.

The advent of large, homogeneous data sets from the Keck I 10-m
telescope obtained with the HIRES spectrograph at high resolution
($R\sim 36,000$) and high signal-to-noise \cite{Hu95},\cite{Kim97} has
made it possible to explore the evolution of the low column density
forest in some detail.  The present analysis uses data from the
spectra of five quasars (Q1623+268, Q1700+643, Q0014+813, Q0302--003,
Q1422+231) which span a redshift range from $z_{em} = 2.521 \to
3.620$, and whose forest clouds have been analyzed over a range from
$z_{abs} = 2.17 \to 3.51$.  This wavelength region was chosen to
sample the `quiet' forest, and to avoid the region of the proximity
effect, and in the case of two partial Lyman limit systems, to avoid
the region within $\pm1000$ km s$^{-1}$ of these systems.  We have
extended the redshift range treated in this work to $z\sim4$ using
the results ($3.40 < z_{abs} < 4.00$) of Lu et al.\ 1996 \cite{Lu96},
who employ a different fitting algorithm but use similar modeling
analysis to correct for selection effects and incompleteness. Details
of the profile fitting can be significant \cite{KT97}, as is discussed
in more detail by Kim et al.\ \cite{Kim97b} in these proceedings,
and a consistent analysis over the full redshift range using a single
procedure should ultimately be carried out, but the Lu et al.\ data
seems fully consistent with the extrapolated trends from $z=2.2 \to 3.5$.

\section{The Differential Density Distribution Function}
The spectrum of HI column densities can be characterized by a power
law with slope $\sim -1.5$ over a wide range of column densities
(Figure~1) but with a marked deficiency of clouds relative to
this fit in the region $10^{14.3} < N_{\rm HI} < 10^{16.3}~{\rm cm}^{-2}$
\cite{Pet93},\cite{Hu95}.

\begin{figure}[h]
\centerline{\vbox{
\psfig{figure=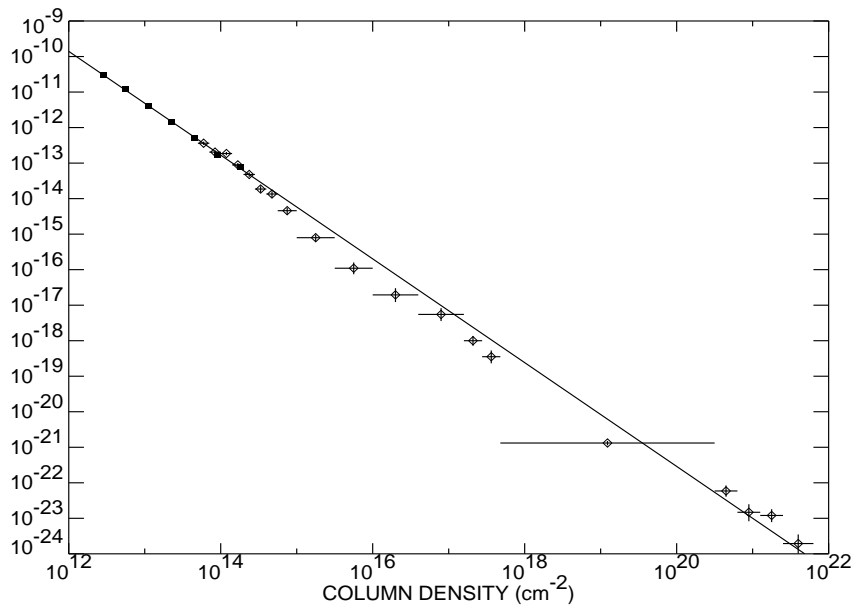,height=8.cm,width=12.cm}
}}
\caption[]{Spectrum of HI column densities at $z\sim3$.  The best power-law 
fit ($\propto N_{\rm HI}^{-1.46}$) to the number of clouds per unit column
density obtained for the weak forest clouds ({\it filled squares\/}) at
$\langle z \rangle = 2.85$ by Hu et al.\ 1995 \cite{Hu95} is shown extrapolated
through the points of Petitjean et al.\ 1993 \cite{Pet93} ({\it diamonds},
with $1\sigma$ error bars) out to column densities of $10^{22}~{\rm cm}^{-2}$
using their notation for this function, and provides a surprisingly good 
description over nearly 10 orders of magnitude in column density.  The
deficiency of clouds noted above $N_{\rm HI} > 10^{14.3}~{\rm cm}^{-2}$ 
is a real effect which cannot be attributed to mis-estimates in the column
density due to saturation effects.}
\end{figure}

\begin{figure}[t]
\centerline{\vbox{
\psfig{figure=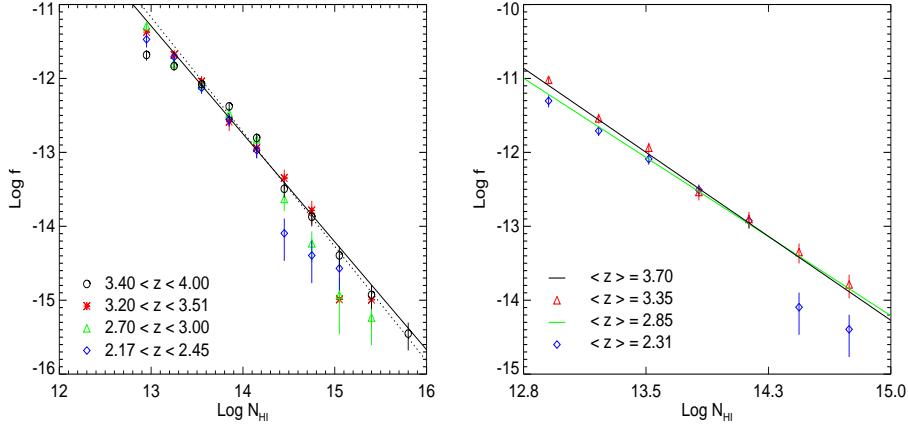,height=5.7cm,width=12.cm}
}}
\caption[]{({\it Left Panel}): The distribution of HI column densities
by redshift bin, with $1\sigma$ error bars.  The solid line shows
the power-law fit ($\propto N_{\rm HI}^{-1.46}$) shown in Figure~1
and the dotted line shows the single power-law fit obtained by Lu
\et \cite{Lu96} for the $\langle z\rangle = 3.7$ data ({\it open
circles}).  At $N_{\rm HI} \le 10^{14}\ {\rm cm}^{-2}$, the distribution
functions at different redshifts look very similar, however,
at $N_{\rm HI} \ge 10^{14}\ {\rm cm}^{-2}$, the deficiency in
the number of forest clouds compared to the expected power-law
distribution depends on redshift.  The point of deviation from the 
power law moves to smaller $N_{\rm HI}$ as redshift decreases,
indicating the rapid evolution in the higher $N_{\rm HI}$ forest clouds.
({\it Right Panel}): Incompleteness corrected differential density
distribution functions compared for $\langle z\rangle = 2.31$ and
$\langle z\rangle = 3.35$ systems.  The reference lines show the
power-law fits of the left-hand panel.
}
\end{figure}
Kim et al.'s 1997 \cite{Kim97} analysis of the differential density distribution
function (DDF) by redshift interval clearly shows an evolution in the departure
from the power-law fit (Figure~2).  In the left-hand panel of Figure~2 the
$\langle z \rangle = 3.70$ data \cite{Lu96} show only a slight steepening
({\it dotted line}) at the higher column densities, but by 
$\langle z \rangle = 3.35$ a deficiency in the number density of clouds
at $N_{\rm HI} > 10^{14.8}~{\rm cm}^{-2}$ can be seen, while clouds
at $\langle z \rangle = 2.85$ and $\langle z \rangle =2.31$ show deviations
from the power-law extrapolation at $N_{\rm HI} > 10^{14.3}~{\rm cm}^{-2}$.
This departure from a power law can be easily seen in the right-hand panel
of Figure~2, where the distribution functions of the $\langle z \rangle = 2.31$
({\it diamonds}) and $\langle z \rangle = 3.35$ ({\it triangles}) absorbers, 
now corrected for incompleteness, are shown.  There are {\it no\/} clouds 
with $N_{\rm HI} > 10^{15.2}~{\rm cm}^{-2}$
at $\langle z \rangle = 2.31$ despite the fact that the redshift path is
1.5 times larger than at $\langle z \rangle = 3.35$, so the deficiency of
high column density clouds in the lower redshift range is not due to 
observational selection.  We conclude that we are seeing a rapid evolution
with redshift in the number density of absorbers at high column density,
with the break in the density distribution function strengthening and
migrating to lower column densities as redshift decreases.

\begin{figure}[t]
\centerline{\vbox{
\psfig{figure=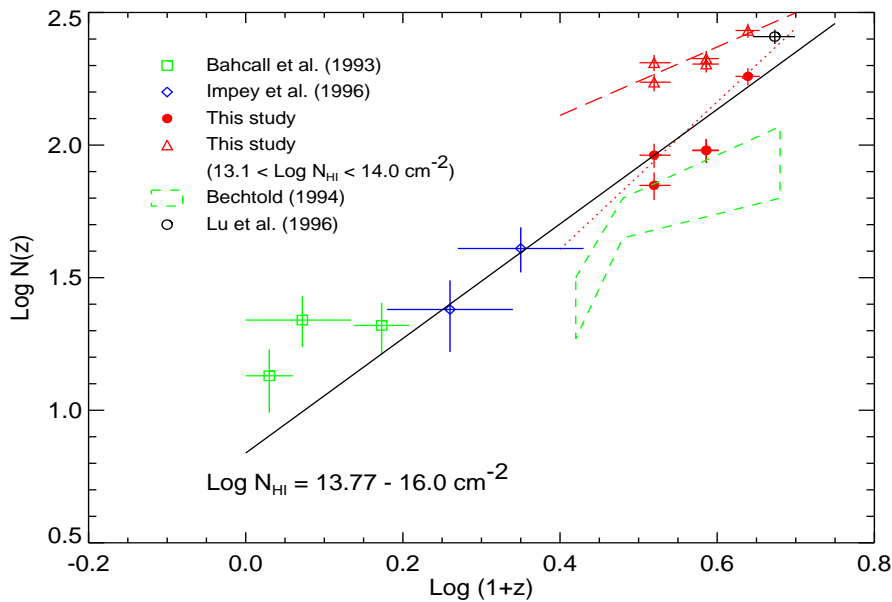,height=8.cm,width=12.cm}
}}
\caption[]{$N(z)$ vs.\ redshift for the present data ({\it filled
circles}) and the HST observations ({\it open squares}; {\it open diamonds}).
The solid line shows the
maximum likelihood fit to the HST data and the high column density
Keck data (slope $\gamma=2.15$).  The dotted line shows the fit to
the Keck data only, with a slope of $\gamma=2.78$.  Triangles show
the forest clouds over just the column density range $N_{\rm HI} =
10^{13.1}-10^{14}\ {\rm cm}^{-2}$, with a long dashed line fit with
$\gamma = 1.19$. The $1\sigma$ error bars and redshift binning are
indicated for each point.}
\end{figure}
Turning to the number density evolution of Ly$\alpha$ clouds with redshift,
where the number density of clouds per unit redshift per line of sight
is expressed as: $N(z) = N_0 (1+z)^{\gamma}$, we can compare the Keck
data on the high redshift systems with the low redshift systems studied
with {\sl HST\/} \cite{Bah93},\cite{Imp96}, while keeping in mind that
$N(z)$ will be sensitive to the adopted threshold $N_{\rm HI}$ (a reflection
of the departure from the single power-law description at $N_{\rm HI} >
10^{14.3}~{\rm cm}^{-2}$).  Figure~3 shows a plot of log$\,N(z)$ vs.
log$\,(1+z)$ giving the maximum likelihood fit to the {\sl HST\/} data and
the high column density Keck data ($\gamma=2.15$).  The adopted column
density range used for the fit, $N_{\rm HI} = 10^{13.77} - 10^{16.0}~{\rm
cm}^{-2}$, corresponds approximately to the conventional equivalent width
threshold of $W=0.32$~\AA\ for comparisons with the lower redshift data
\cite{Bah93},\cite{Imp96},\cite{Bec94}.  By contrast, the best fit to the
slope of $N(z)$ for the low column density systems in the Keck data is
$\gamma = 1.19$.  The definition of $N(z)$ yields $\gamma=1$ for $q_0=0$
and $\gamma=0.5$ for $q_0=0.5$ in the case of non-evolving clouds, and the
slopes for the data in different column density ranges in Figure~3 can be
understood in terms of the evolution of the break in the DDF: the number
density of the higher column density systems ($\gamma=2.15$) is evolving
rapidly, while the number density of the lower column density systems
($10^{13.1} < N_{\rm HI} < 10^{14.0}~{\rm cm}^{-2}$; $\gamma=1.19$) is not.

\section{The Evolution of {\it b}-Values}
A second interesting issue is the evolution of line-widths, characterized
by the Doppler parameter $b$ obtained from profile fitting, 
whose behavior reflects the thermal/ionization history of the IGM \cite{Hae97}.
Hu \et 1995 \cite{Hu95} found a typical mean $b$ of 28 km~s$^{-1}$
with a minimal $b_c$ = 20 km~s$^{-1}$ from the analysis of 
$\langle z\rangle = 2.85$ systems. At $\langle z\rangle = 3.70$
Lu \et \cite{Lu96} report a mean $b$ of 23 km~s$^{-1}$ and
cutoff $b_c$ = 15 km~s$^{-1}$, showing a trend to narrower
absorbing components at higher redshift.

In order to quantify these descriptions, it is important to run model
simulations to understand the nature of blending, selection, and
incompleteness effects on the analysis of the observed data.  The
procedure used \cite{Hu95},\cite{Lu96} is to input a truncated Gaussian
distribution for the Doppler parameter values, generate a series of
simulated artifical spectra, and then apply the standard profile-fitting
methods to match the recovered column densities and $b$ values to the
input spectra and to the observations. 
\begin{figure}[t]
\centerline{\vbox{
\psfig{figure=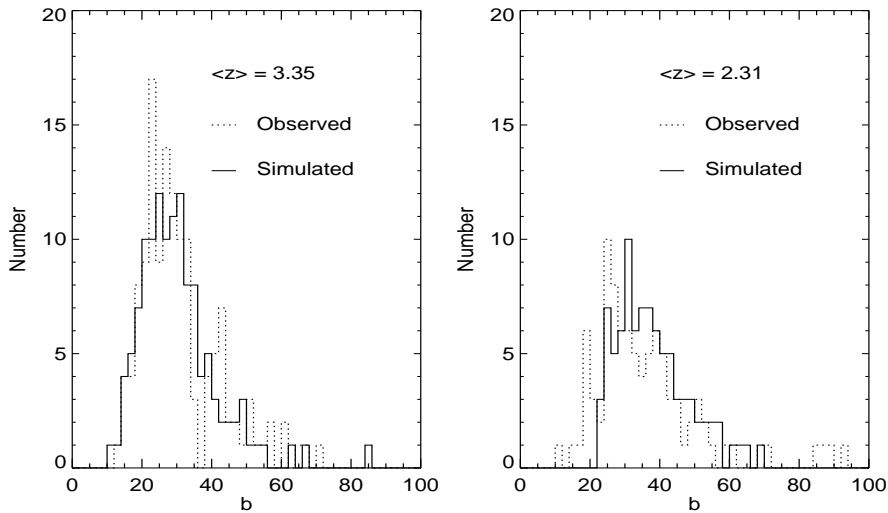,height=6.6cm,width=12.cm}
}}
\caption[]{
Observed (dotted histogram) and simulated (solid histogram) $b$
distributions. For the $\langle z \rangle = 3.35$ clouds, the low
cut-off $b$ value is $\sim17~{\rm km\ s}^{-1}$. For the $\langle
z \rangle = 2.31$ clouds, there is no single best-fit assumed
distribution for $b$. Assuming that most narrow lines are
unidentified metal lines towards Q1700+643 and ignoring the lines
with $b < 22~{\rm km\ s}^{-1}$, the cut-off $b$ of $24~{\rm
km\ s}^{-1}$ which is shown in the figure best fits the data.
}
\end{figure}

Both the cutoff and median $b$ values appear to increase with decreasing
redshift --- the cutoff $b$ value's increase from 15 km~s$^{-1}$ at
$\langle z \rangle = 3.70$, to $17~{\rm km\ s}^{-1}$ at 
$\langle z \rangle = 3.35$, to $22~{\rm km\ s}^{-1}$ at 
$\langle z \rangle = 2.85$, to $24~{\rm km\ s}^{-1}$ at 
$\langle z \rangle = 2.3$ presumably reflecting an increase in the
turbulent and thermal broadening over this period.

\section{Summary}
In broad outline the cloud evolution described here probably follows
naturally from the decreasing IGM density with decreasing redshift,
which drives down the column density at which shocking occurs
\cite{Mue97}, and the increasing level of structure which results in
more turbulent components.  However, reproducing the results in detail
will require full comparison with a wide suite of numerical models,
which may provide useful constraints on the history of the ionization
and heating of the IGM.

\acknowledgements{This work was partly supported by NSF grant AST 96-17216,
and used observations obtained with Steve Vogt's HIRES spectrograph for the 
Keck I telescope.  The W.~M.\ Keck Observatory is operated as a scientific 
partnership between the California Institute of Technology, the University
of California, and the National Aeronautics and Space Administration.}


\vfill
\end{document}